\documentclass[aps,aip,reprint,floats]{revtex4-1}
\usepackage{amsmath,amsfonts,graphicx,color}

\begin{document}
\bibliographystyle{revtex}

\textwidth 16cm \textheight 23cm \topmargin -1cm \oddsidemargin
0cm \evensidemargin 0cm

\title{Wave-particle interactions with parallel whistler waves: nonlinear and time-dependent effects revealed by Particle-in-Cell simulations}
\author{Enrico Camporeale$^{a}$}\email{e.camporeale@cwi.nl}
\author{Gaetano Zimbardo$^{b}$}

\affiliation{a) Center for Mathematics and Computer Science (CWI), 1098 XG, Amsterdam, Netherlands} 
\affiliation{b) Department of Physics, University of Calabria, Ponte P. Bucci, Cubo 31C, I-87036 Rende, Italy}

\date{\today}

\begin{abstract}
We present a self-consistent Particle-in-Cell simulation of the resonant interactions between
anisotropic energetic electrons and a population of whistler waves, with parameters relevant
to the Earth’s radiation belt. By tracking PIC particles, and comparing with test-particle
simulations we emphasize the importance of including nonlinear effects and time evolution in
the modeling of wave-particle interactions, which are excluded in the resonant limit of quasi-
linear theory routinely used in radiation belt studies. In particular we show that pitch angle
diffusion is enhanced during the linear growth phase, and it rapidly saturates { well before a single bounce period}. This calls into question the widely used bounce average performed in most radiation belt diffusion calculations. Furthermore, we discuss how
the saturation is related to the fact that the domain in which the particles’ pitch angle diffuse
is bounded, and to the well-known problem of $90^\circ$ diffusion barrier.
\end{abstract}


\maketitle


%
%

\section{Introduction}
Resonant wave-particle interactions represent one of the most important mechanism that regulate the scattering and loss of energetic particles in the radiation belts \cite{thorne10}.
Among the different waves that can be generated and propagate in the Earth's radiation belt,
much attention has been devoted to whistler waves. They are right-handed polarized electromagnetic waves
with frequencies ranging between the ion and electron gyrofrequency.
Whistler waves are associated with so-called chorus modes that usually present two characteristic frequency bands with a 
gap in between. An interpretation of such a gap based on linear theory has recently been presented in \cite{fu14}.
Whistler waves can be generated by a kinetic instability driven by a temperature anisotropy (with the temperature
in the perpendicular direction greater than in the parallel direction).
For instance, equatorial whistler-mode chorus can be excited by cyclotron resonance 
with anisotropic 10-100 KeV electrons injected from the plasma sheet
\cite{summers07,jordanova10}.
The generation and propagation of whistler-mode chorus in the Earth's radiation belt has been intensively studied:
simulations of rising tones have been performed in \cite{katoh07a, katoh07b, hikishima09} for loss-cone distributions 
and in \cite{tao14} for bi-Maxwellian; the non-linear wave growth mechanism has been studied, e.g., in \cite{omura08, summers12, omura12,gao14b}.\\
The current paradigm for modeling wave-particle interactions in the radiation belt is based on kinetic quasi-linear theory. 
It means that the particle distribution function follows a diffusive scattering in the adiabatic invariants space, assuming a broadband wave spectrum and low amplitude fluctuations \cite{kennel66, jokipii66, lyons73}.
Although quasi-linear theory was primarily developed to study the saturation of a linear instability, due to the distribution
function diffusion in phase space, and the formation of a plateau, it is customary to employ the so-called 'resonant limit' in
radiation belt simulations, that is the limit in which the linear growth rate tends to zero, and the distribution reaches a marginally stable state \cite[e.g.,][]{summers98}. The underlying assumption of such approach is that diffusion due to a quasi-stationary wave spectrum (and the associated longer timescale) is more relevant than the short-lived diffusion occurring during the linear phase of the instability. Alternatively, one can
justify the use of the resonant limit by assuming that the diffusing particles are unrelated to the wave, in the sense that they do not belong to the part of the distribution function that is responsible for the development of the kinetic instability.
It is important to remind that the calculation of the diffusion coefficients usually incorporate 
the interactions between waves and resonant particles only, assuming a stationary wave spectrum with a given amplitude \cite{summers05, glauert05, summers07}. Not only this is not realistic during the linear growth phase of the instability, but we note that the dominance of resonant over non resonant interactions in a turbulent wave field has been recently questioned by \cite{ragot12}.
{ It is the goal of this paper to quantify the effect of a non-stationary wave spectrum during the linear phase of the instability, and the corresponding discrepancy in pitch-angle diffusion rate, with respect to the widely used resonant limit approximation}.  
This is achieved by performing Particle-in-Cell (PIC) simulations, that is by generating the instability in a completely self-consistent way, 
 without the need of further assumptions (contrary to quasi-linear theory). The PIC simulations results will be compared with test-particle
 simulations, { where a stationary whistler wave spectrum is assumed}. Such comparison will emphasize the inadequacy of employing the resonant limit of quasi-linear theory for the case studied.
 Indeed, by tracking resonant particles, we show that pitch angle scattering is tremendously enhanced
 during the linear growth phase of the instability. The decrease of anisotropy due to the development of the whistler 
 instability results in a rapid precipitation in the loss-cone, in a measure much larger than predicted by quasi-linear diffusion
 due to a non-growing wave activity.\\

Although the distribution function diffusion
proceeds in a three-dimensional space (for instance, in energy, pitch angle, and radial directions), the timescale between energy/pitch angle and radial diffusion is very well separated \cite{camporeale13}, and hence in this paper we focus on two-dimensional diffusion only.

When electron pitch angle undergoes a stochastic quasi-linear diffusion, its mean squared displacement $\langle\Delta \alpha^2\rangle$ 
 is expected to grow linearly in time.
 Indeed, for normal diffusion, the quasi-linear diffusion coefficient $D$ and $\langle\Delta \alpha^2\rangle$  
 are related by the celebrated
 Einstein relation $\langle\Delta \alpha^2\rangle = 2D t$, at least for a certain time range.
 In this regime, \cite{tao11} have shown that test-particle simulations are an excellent tool to test the validity of quasi-linear diffusion,
 by simply assessing the Einstein relation, with the pitch angle statistics gathered from the simulation particles.\\
Diffusion in pitch angle presents a crucial difference with standard diffusion in physical space, which is commonly represented and understood in statistical terms with the concept of random walk. 
The difference is that the pitch angle coordinate is defined on a bounded domain, that is $\alpha\in[0^\circ,180^\circ]$.
It is well known that diffusion on bounded domains produce a mean squared displacement that saturates in time (in the case of homogeneous
diffusion coefficient, to a value that is proportional to the domain length), \cite[e.g.,][]{metzler00, bickel07}.
Interestingly, such important feature of pitch angle scattering has not been emphasized and analyzed in the radiation belt literature.
Of course, once the distribution in pitch angle becomes close to a saturated state, the Einstein relation becomes meaningless, 
because although individual particles continue diffusing, the diffusion coefficient cannot be correctly evaluated by means 
of the mean squared displacement (which becomes constant in time). An important consideration, in a space weather perspective, 
is whether such (local) saturation of the pitch angle distribution can occur on a time scale which is much shorter than the bounce period, over which the diffusion equation is usually averaged (in order to reduce the dimensionality of the problem, and simplify the calculation of the 
diffusion coefficients).\\
Another important aspect of quasi-linear diffusion that apparently has been overlooked in radiation belt studies is the so-called '$90^\circ$ problem', that, on the other hand, has been the focus of several works in cosmic-ray acceleration context (see, e.g. \cite{shalchi05, tautz08, qin09}). In such context, it has been shown that quasi-linear theory underestimates the diffusion through $90^\circ$ angle, which represents 
an effective diffusion barrier in pitch angle space (we do not mean, by barrier, total reflection, but a very small diffusion). 
Diffusion through $90^\circ$ is instead sizable for some particles (depending on their
energy), and this can be incorporated in a diffusion model when considering second-order and nonlinear effects.\\

The importance of a correct assessment of the particle diffusion stems from the fact that particle lifetime is  
approximately estimated, in the weak diffusion limit, as the inverse of the diffusion coefficients evaluated at the equatorial loss-cone angle \cite{shprits07,albert09, mourenas12}, by assuming that the scattering remains diffusive through multiple bounce periods \cite{summers07b}.
As a reference number, we note that the bounce period at $L=4.5$ for a 1 MeV electron traveling with loss-cone
 equatorial pitch angle is approximately 0.24 seconds, or $1.4\cdot10^4 \Omega_e^{-1}$ ($\Omega_e$ being the equatorial electron gyrofrequency).
 An important open question then is whether pitch angle scattering retains its diffusive character for several bounce periods.
As we will show, the whistler instability (for sufficiently large temperature anisotropy) saturates within few hundreds electron gyroperiods,
and the local distribution reaches a quasi-stationary equilibrium, by the time the instability saturates.
 
\section{Methodology}
We present a one-dimensional Particle-in-Cell simulation performed with the 
implicit moment-method code PARSEK2D \cite{markidis09,markidis10}. 
We are concerned here with a typical situation in the equatorial region of the radiation belt,
where whistler waves can be excited by a tenuous population of hot anisotropic electrons.
The homogeneous background magnetic field is $B_0 = 4\cdot10^{-7} T$, 
corresponding approximately to the Earth's equatorial value at $L\sim 4.3$, and is aligned with the box.
We neglect the dipolar nature of the Earth's magnetic field, and therefore our conclusions are valid only for small regions close to the equator.
On the other hand, the timescales investigated are much shorter than the bounce period, and we have estimated that the most energetic particles in the
simulation would experience, in a dipole field, a change of less than $5\%$ in magnetic field amplitude. 

The electron population has a density of 15 cm$^{-3}$, and it is composed for $98.5\%$
by a cold isotropic Maxwellian (1 eV), and for $1.5\%$ by an anisotropic relativistic bi-Maxwellian 
distribution $f(v_{||},v_\perp) \propto \exp\left[-\alpha_\perp \gamma-(\alpha_{||} - \alpha_\perp)\gamma_{||}\right]$ 
(with $\gamma = (1-v^2/c^2)^{-1/2}$, $\gamma_{||} = (1-v_{||}^2/c^2)^{-1/2}$, $c$ the speed of light, and
parallel and perpendicular refer to the background magnetic field) \cite{naito13,davidson89}.
We choose $\alpha_{||}=25$, and $\alpha_\perp=4$. The hot population velocity distribution function
has standard deviations $\sqrt{\langle v_{||}^2 \rangle} = 0.175$, and $\sqrt{\langle v_\perp^2 \rangle} = 0.325$ (normalized 
to speed of light)
corresponding to nominal temperatures of 8 KeV and 30 KeV, respectively.
Thus, the initial anisotropy of the suprathermal component is $T_\perp/T_{||}= 3.75$.
The ratio between electron plasma and gyro-frequency is $\omega_p/\Omega_e\simeq 3.1$.
The box length is $L=400 c/\Omega_e$. We use 8,000 grid points, a timestep $\Delta t\Omega_e=0.015$,
and 500 particles per cell (excluding tracking particles, see later).
Since the hot population drives the instability, we use 400 and 100 particles per cell for the cold and for the hot population, respectively,
 resulting in a higher resolution in velocity for the latter. The ions are immobile.

\begin{figure}
 \includegraphics[width=15 cm, height=10 cm]{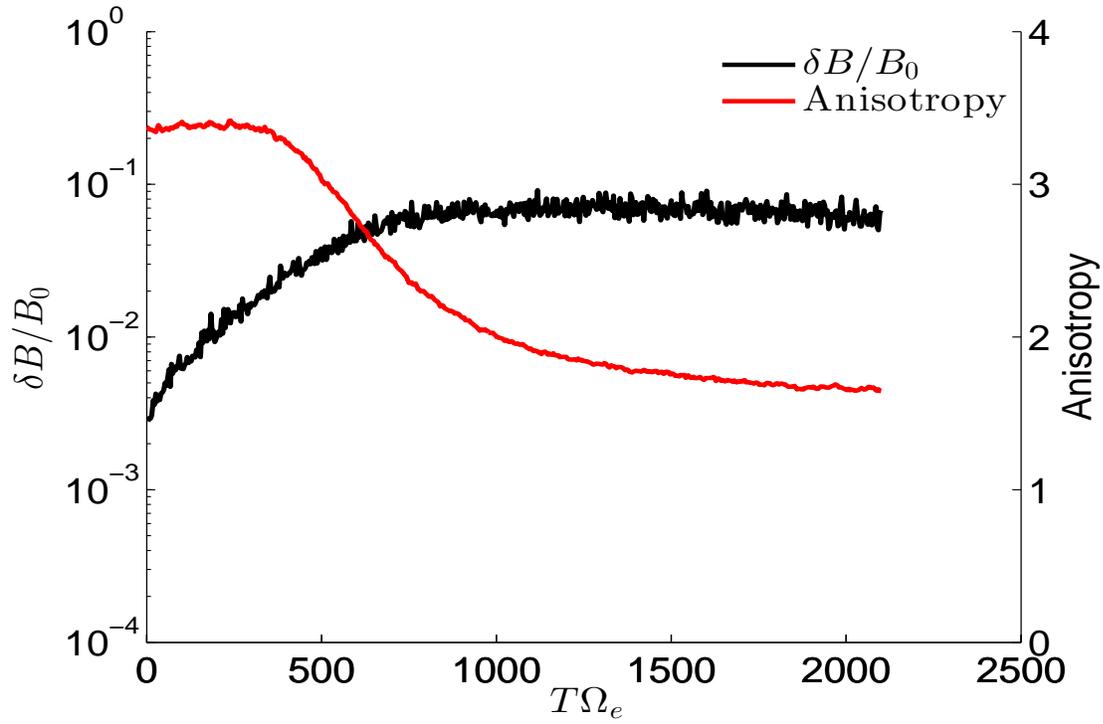}%
 \caption{PIC simulation results. Magnetic field relative amplitude $\delta B/B_0$ (black line, left axes in logarithmic scale), and anisotropy $T_\perp
/T_{||}$ (red line, right axes in linear scale) as a function of time $T\Omega_e$}
 \label{fig:anisotropy}
 \end{figure}

\begin{figure}
 \includegraphics[width=15 cm, height=10 cm]{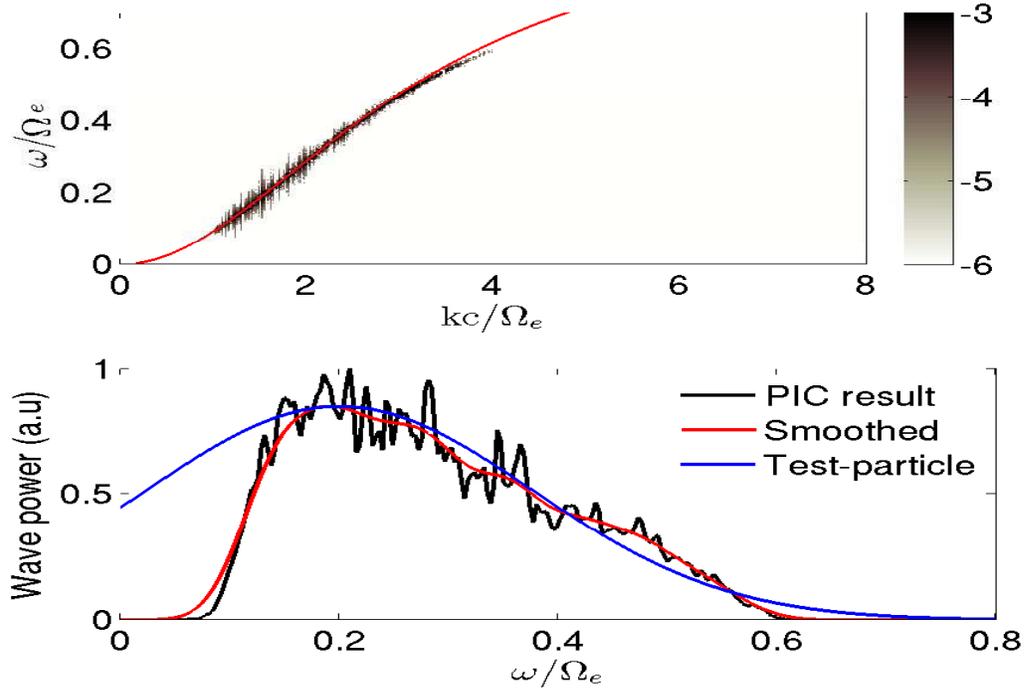}%
 \caption{Top: spectrogram of magnetic fluctuations in logarithmic scale (frequency vs wavevector). The black dots indicate PIC simulation results.
 The red line shows the
 dispersion relation from cold plasma theory. Bottom: wavepower as function of frequency. Black, red, and blue line
 represent the result from PIC simulations, a smoothed fit, and the Gaussian spectrum used for test-particle simulations.}
 \label{fig:spectrogram}
 \end{figure}

Temperature anisotropy instabilities have a 'self-destructing' character, in the sense
that the generated electromagnetic fluctuations reduce the anisotropy that drives the instability, and therefore a marginal
stability condition is usually rapidly reached \cite{lu04,camporeale08,gary14,hellinger14,lu10}.
Figure \ref{fig:anisotropy} shows the reduction of temperature anisotropy (red line, right axes) and
the increasing magnetic field amplitude (black line, left axes). The linear instability saturates around 
the time $T\Omega_e=900$, and although the anisotropy response to the magnetic field fluctuations is somewhat delayed,
the correlation is clear. Indeed, in the early linear phase the instability grows starting from small values, without affecting the electron distribution function until $\delta B/B_0 \simeq 0.02$ is attained.\\
In Figure \ref{fig:spectrogram} we show the numerical dispersion relation (magnetic field amplitude, top panel), and the wavepower as function of 
frequency (bottom panel), calculated over the entire simulation time
$T\Omega_e=2100$.
The red line in the top panel represents the whistler dispersion relation derived from cold plasma theory,
which, despite neglecting the suprathermal component, is still a good approximation.
Note that the wavepower is peaked around $\omega/\Omega_e\sim0.2$ and it is confined within $\omega/\Omega_e<0.6$.
In the bottom panel the black line denotes results from the PIC simulation (and therefore are relatively noisy). The 
red line is a smoothed fit of the PIC result, and the blue line is the { stationary} spectrum that will be used for the test-particle calculations.
It is a Gaussian centered in $0.2\Omega_e$ with width equal to $0.25\Omega_e$. Although it overestimates the 
wavepower at small frequencies, this is a good approximation of the PIC results in the range $[0.2-0.6]\Omega_e$.\\
\subsection{Tracking particles}
We emphasize that the PIC approach is first-principle and does not rely on any of the assumptions employed for quasi-linear diffusion codes
or test-particle simulations. Moreover, the diagnostics on the particle scattering is readily available.
In the simulations presented in this paper we have tracked 8 groups composed of 32,000 particles each, that have been 
initialized with different pitch angle and energies.
Specifically, the initial pitch angles range from $10^\circ$ to $80^\circ$, in intervals of $10^\circ$. 
The initial velocity is chosen such that the particles satisfy (at initial time) the resonant condition
\begin{equation}\label{resonance}
 \omega- k v \cos\alpha = \Omega_e\sqrt{1-v^2/c^2}.
\end{equation}
for a chosen value of $\omega$. 
The wavevector $k$ in Eq. (\ref{resonance}) is derived from the cold plasma dispersion relation 
for whistlers:
\begin{equation}
 \left(kc/\omega\right)^2 = {1-\frac{\omega_p^2/\Omega_e^2}{\omega(\omega-\Omega_e)}}.
\end{equation}

For each value of initial pitch angle the tracking particles are initially resonant with $\omega/\Omega_e=0.3$.
The initial energy for each group of particles is summarized in Table 1.
Note that for particles with $\alpha<90^\circ$, the resonant wave is counter-propagating, that is $k<0$.
The tracked particles are initially uniformly distributed in the box.

\subsection{Test-particle simulations}
Although the main focus of this paper is to comment results derived from PIC simulations, we are also interested in comparing the results against test-particle simulations. 
The interest for test-particle methods in the radiation belt studies stems, on one hand, from their computational speed,
and on the other hand from the relationship that exists with quasi-linear diffusion codes, { in the resonant limit approximation}.
Indeed, it is expected that when the assumptions of quasi-linear theory are satisfied, the Einstein relation
between quasi-linear diffusion coefficients and test-particle mean squared displacements
$\langle\Delta \alpha^2\rangle = 2D t$
holds.
Ref. \cite{tao11} have successfully shown that this is the case for small-amplitude parallel propagating whistler, 
and they have later proved the breakdown of quasi-linear theory for larger amplitudes \cite{tao12}.
See also \cite{liu10} for a discussion on the departure time, that is the time at which $\langle\Delta \alpha^2\rangle$ departs from the 
Einstein relation.\\
The advantage of test-particle simulations is that one can specify the electromagnetic field at 
any spatial location with any desired accuracy. Of course, this is in contrast to gridded methods such as PIC where the field 
must be interpolated from the grid to the particle locations. On the other hand,
test-particle codes lack the self-consistency and conservation properties of PIC (for instance, particles can be indefinitely accelerated).
\\
In this paper we use the same code described in \cite{tao11}. { Test particles are moved on a prescribed, stationary, whistler wave spectrum. The wave spectrum is approximated with the superposition of 200 cold plasma modes equally spaced
in frequency between $\omega/\Omega_e = 0.009$ and $\omega/\Omega_e = 0.6$}, each of them weighted according to 
the Gaussian curve shown in Figure \ref{fig:spectrogram} (bottom panel, blue curve).
For each run the statistics is performed on 400 particles, 
advanced in time with a timestep $\Delta t\Omega_e=0.01$.

\section{Results}
\begin{figure}
 \includegraphics[width=15 cm, height=10 cm]{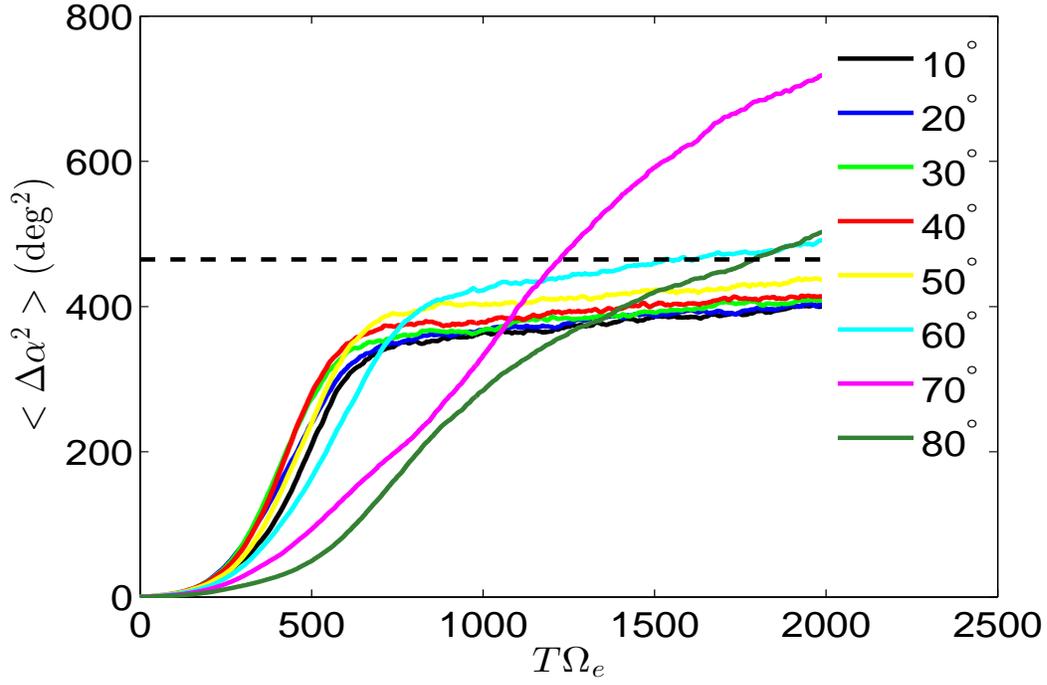}%
 \caption{Evolution of the pitch angle mean squared displacement $\langle \Delta\alpha^2\rangle$ in time for tracked particles. Different colors are for different
 initial pitch angle. The black dashed line denotes the saturation value 465 deg$^2$ (see text for discussion).}
 \label{fig:variance_pa}
 \end{figure}

The main diagnostics that we study is the mean squared displacements in pitch angle and energy.
Such quantities are denoted as $\langle\Delta \alpha^2\rangle$,  $\langle\Delta E^2\rangle$, and $\langle\Delta E \Delta \alpha\rangle$
(the mixed diffusion term), where $\Delta\alpha = \alpha-\langle\alpha\rangle$, $\Delta E = E -\langle E\rangle$, and 
$\langle\ldots\rangle$ denotes the average over the whole sample.\\
Figure \ref{fig:variance_pa} shows the development of $\langle\Delta \alpha^2\rangle$ in time for the tracked { PIC} particles. 
The different colors are labeled in the legend and correspond to different initial pitch angles.
An important feature, and one of the main results of the paper, is that for all angles less than $70^\circ$,
the pitch angle mean squared displacement $\langle\Delta \alpha^2\rangle$ shows two distinct
phases: a rapid growth for $T\Omega_e\lesssim 700$, and a much slower growth at later times. This behavior is
nicely correlated with the linear growth phase shown in Figure \ref{fig:anisotropy}.
Moreover, for the simulation time presented, the dashed line represents an 
asymptotic value for the resonant particles with initial pitch angle less than $60^\circ$.
Such dashed line corresponds to the pitch angle variance of an isotropic velocity distribution, but with all the particles bounded
to the $\alpha<90^\circ$ interval. This is simply calculated by defining the particle distribution function $f(\alpha)=\sin\alpha$ 
for $\alpha\leq90^\circ$, and $f(\alpha)=0$ for $\alpha>90^\circ$. The mean value of such distribution is equal to 1 rad = $57.3^\circ$.
The variance in degrees is then calculated as
\begin{equation}
 \left(\frac{180^\circ}{\pi}\right)^2 \int_0^\frac{\pi}{2} (\alpha-1)^2 f(\alpha) d\alpha \simeq 465 \textrm{ (deg$^2$)}.
\end{equation}

\begin{figure}
 \includegraphics[width=15 cm, height=10 cm]{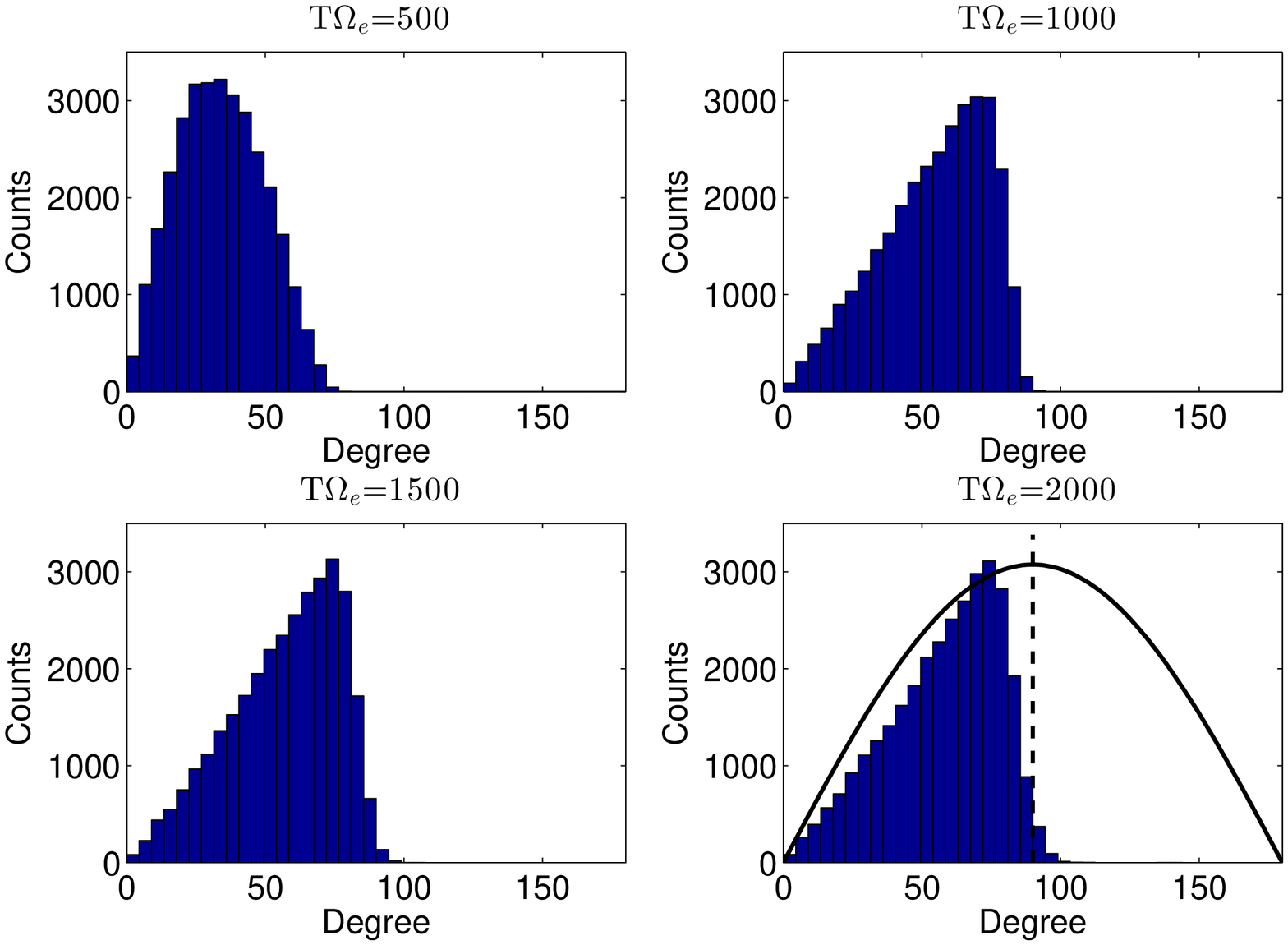}%
 \caption{Histograms of pitch angle distribution for tracked particles with initial pitch angle $\alpha=20^\circ$, at times $T\Omega_e=500$, 1000, 1500, 2000. The solid line in the right-bottom panel represents the isotropic distribution $f(\alpha)=\sin\alpha$, and the vertical dashed line denotes $\alpha=90^\circ$.}
 \label{fig:histo_20_res}
 \end{figure}

\begin{figure}
 \includegraphics[width=15 cm, height=10 cm]{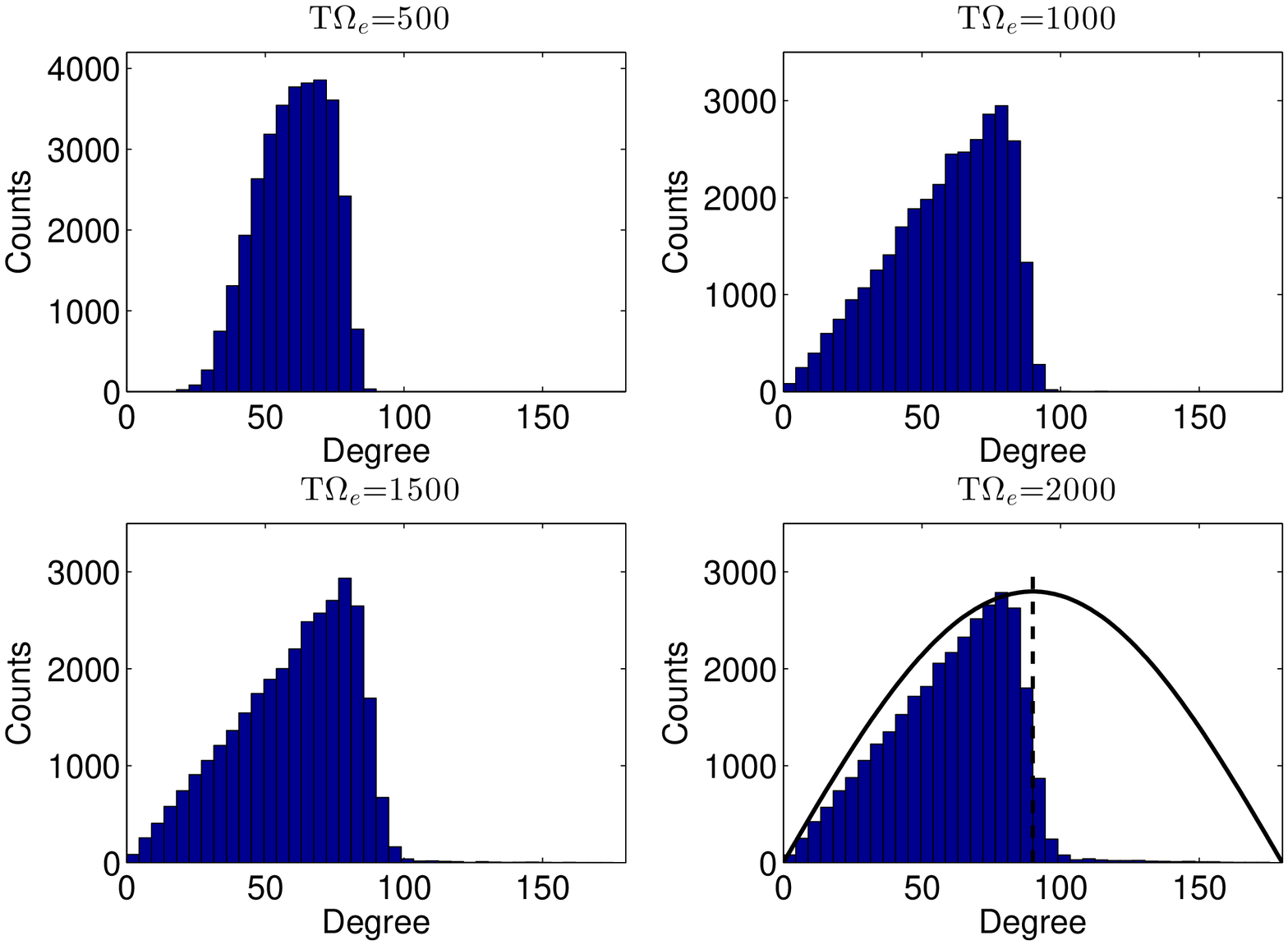}%
 \caption{Histograms of pitch angle distribution for tracked particles with initial pitch angle $\alpha=60^\circ$, at times $T\Omega_e=500$, 1000, 1500, 2000. The solid line in the right-bottom panel represents the isotropic distribution $f(\alpha)=\sin\alpha$, and the vertical dashed line denotes $\alpha=90^\circ$.}
 \label{fig:histo_60_res}
 \end{figure}

\begin{figure}
 \includegraphics[width=15 cm, height=10 cm]{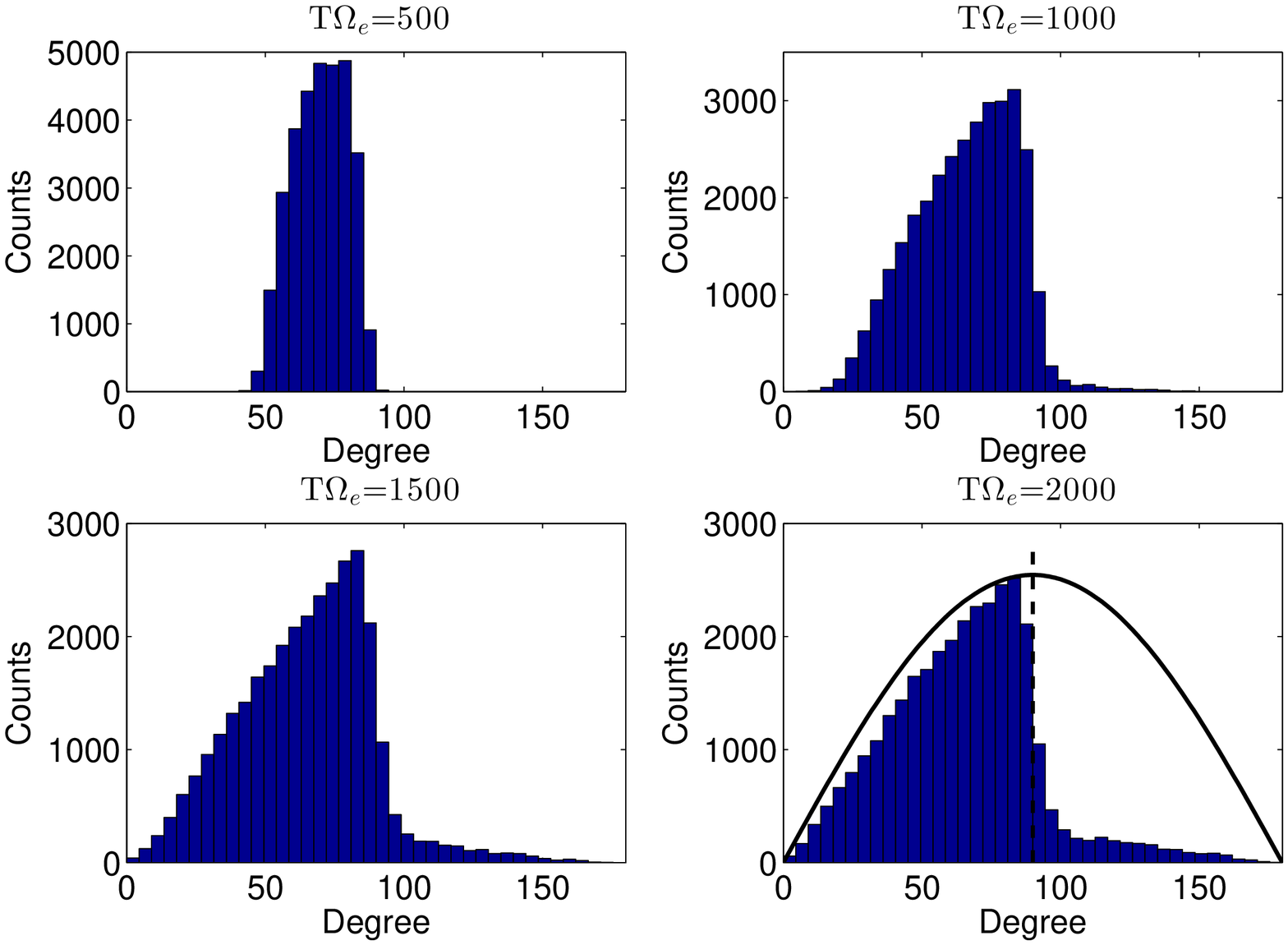}%
 \caption{Histograms of pitch angle distribution for tracked particles with initial pitch angle $\alpha=70^\circ$, at times $T\Omega_e=500$, 1000, 1500, 2000. The solid line in the right-bottom panel represents the isotropic distribution $f(\alpha)=\sin\alpha$, and the vertical dashed line denotes $\alpha=90^\circ$.}
 \label{fig:histo_70_res}
 \end{figure}

\begin{figure}
 \includegraphics[width=15 cm, height=10 cm]{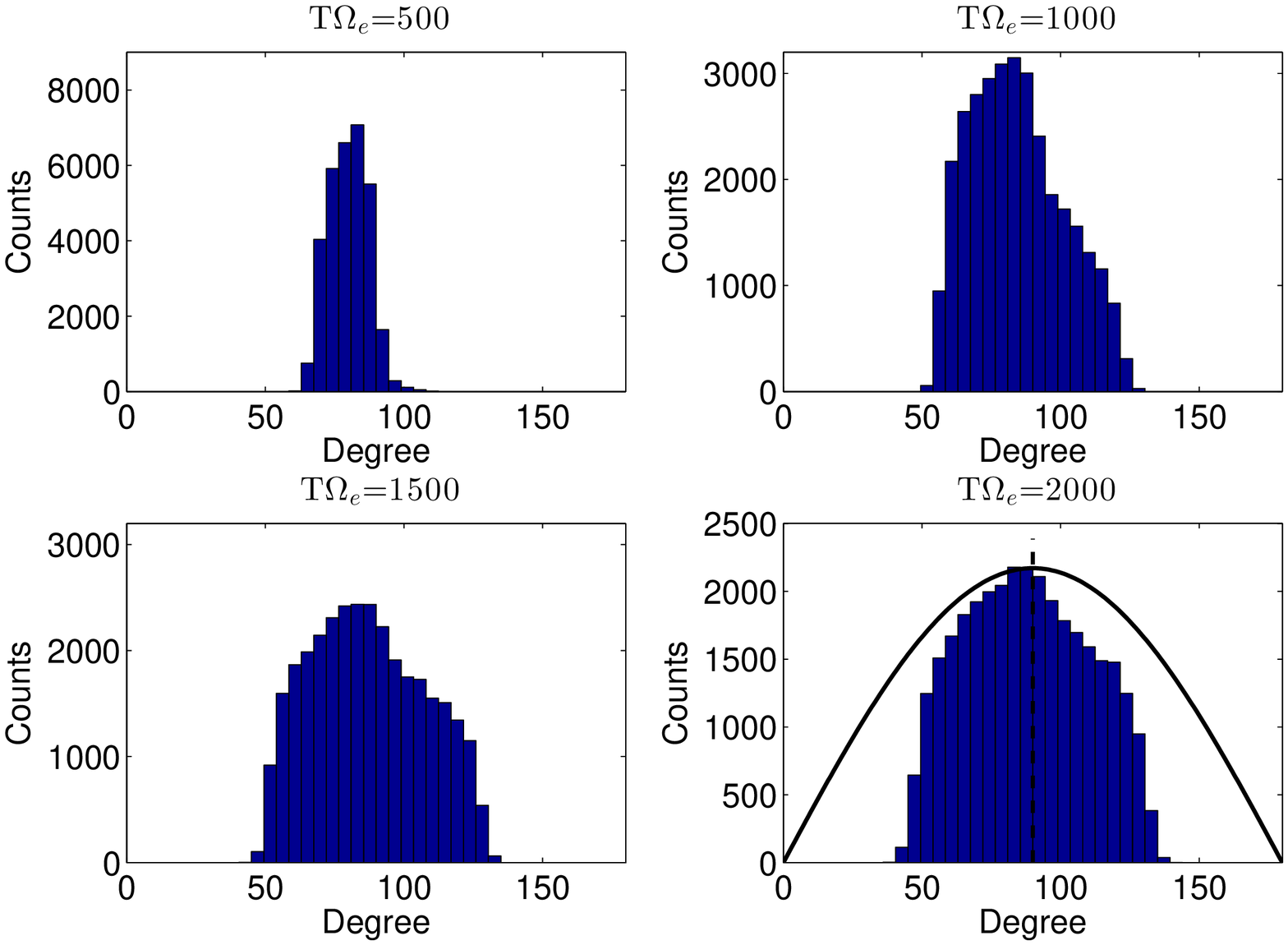}%
 \caption{Histograms of pitch angle distribution for tracked particles with initial pitch angle $\alpha=80^\circ$, at times $T\Omega_e=500$, 1000, 1500, 2000. The solid line in the right-bottom panel represents the isotropic distribution $f(\alpha)=\sin\alpha$, and the vertical dashed line denotes $\alpha=90^\circ$.}
 \label{fig:histo_80_res}
 \end{figure}

\begin{figure}
 \includegraphics[width=15 cm, height=10 cm]{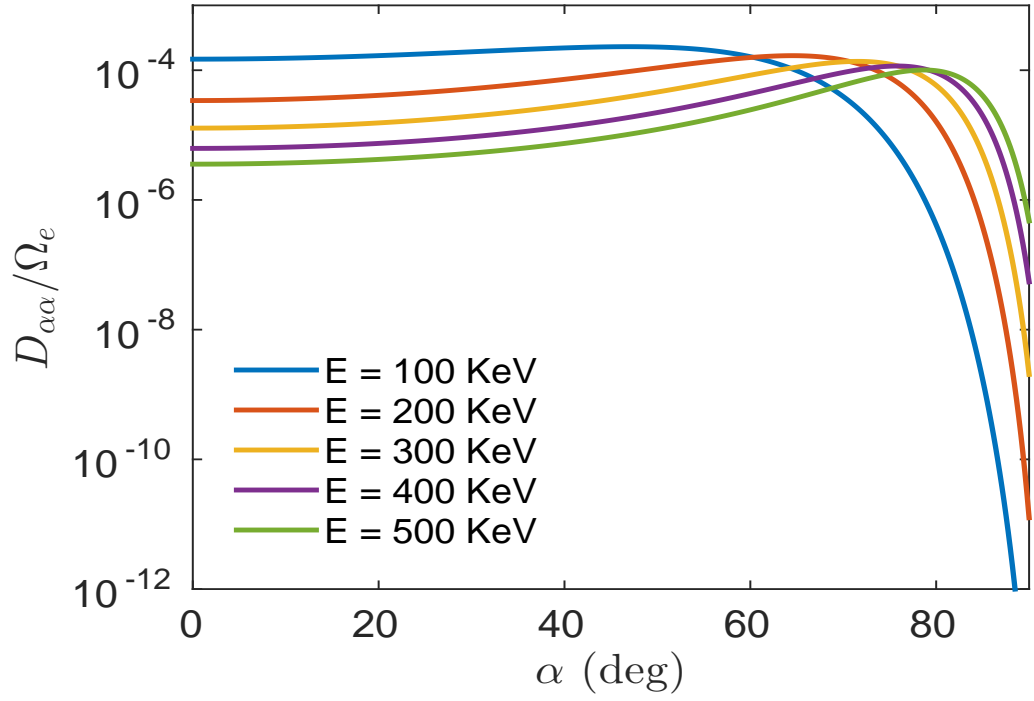}%
 \caption{Diffusion coefficient $D_{\alpha\alpha}$ calculated as function of $\alpha$, for $\delta B/B_0=0.01$, for different energies, ranging from 100 to 500 KeV.}
 \label{fig:Daa_90deg}
 \end{figure}

In order to understand the behavior shown in Figure \ref{fig:variance_pa}, and how the asymptotic value of $465^\circ$ comes about,
we look at the evolution of the distribution in pitch angle at different times.
Figures \ref{fig:histo_20_res}, \ref{fig:histo_60_res}, \ref{fig:histo_70_res}, and \ref{fig:histo_80_res}
show the histograms of the number of { tracked PIC} particles at different angles for times $T\Omega_e = 500, 1000, 1500, 2000$, 
for initial pitch angles $\alpha=20^\circ, 60^\circ, 70^\circ, 80^\circ$, respectively.
We note that $T\Omega_e = 2000$ is still much less than the electron bounce period in the Earth's dipole field, which is estimated at 14,000 $\Omega_e^{-1}$
A common feature of Figures \ref{fig:histo_20_res} and \ref{fig:histo_60_res} (i.e. for initial $\alpha=20^\circ$ and $ 60^\circ$)
is that $90^\circ$ represents a diffusion 'barrier', in the sense that diffusion through $90^\circ$ is very limited, although not exactly null.
The same feature appears for particles with initial pitch angle $\alpha=30^\circ, 40^\circ, 50^\circ$ (not shown).
This is consistent with standard quasi-linear theory which predicts a very small diffusion coefficient at $90^\circ$.
This is shown in Figure \ref{fig:Daa_90deg}, where the Summers coefficient $D_{\alpha\alpha}$  \citep{summers07}
is plotted { as function of pitch-angle for different energies} for a wave amplitude $\delta B/B_0=0.01$ (note that the coefficient scales linearly with the square
of the wave amplitude). For the range of energies and the timescale considered here the pitch
angle diffusion coefficient at $90^\circ$ is essentially null.
We note however that, as \cite{summers07} clarifies, nonlinear effects and phase trapping are not included in the quasi-linear treatment.
The bottom-right panels of Figures \ref{fig:histo_20_res} and \ref{fig:histo_60_res}
also show the analytical isotropic distribution $f(\alpha)$ discussed previously, as a black line, and they support
the argument that since diffusion tends to fill the left half of the distribution, the variance approaches in the time the value of $465$ 
(deg$^2$), as shown in Figure \ref{fig:variance_pa}.
Figure \ref{fig:histo_70_res} has the same format of Figures \ref{fig:histo_20_res} and \ref{fig:histo_60_res},
but now for initial $\alpha=70^\circ$.
The behavior is not very dissimilar, but one can notice a non-negligible fraction of particles diffusing through the $90^\circ$ barrier.
Finally, in Figure 7, we show the histograms for initial pitch angle $\alpha=80^\circ$.
The behavior is now qualitatively different, and this was already evident from the mean squared displacement shown in Figure \ref{fig:variance_pa}.
There is no sign of a diffusion barrier at $90^\circ$, and at the final stage
the distribution is almost symmetrical around $90^\circ$. { This behavior is in strong contrast with the prediction of quasi-linear theory.}
The qualitative difference in pitch angle diffusion across $90^{\circ}$ which is found for particles injected with $\alpha=20^{\circ}$ and $\alpha=70^{\circ}$--$80^{\circ}$ is due to the different initial velocity of resonant particles, which is shown in Table 1 (in terms of energy). Indeed, in order to keep satisfying the resonance condition Eq. (1), when $\alpha$ becomes close to $90^{\circ}$, a larger particle velocity would be required, or, equivalently, a larger wave frequency (remember that $k<0$). This is most critical for particles injected with $\alpha=20^{\circ}$ since the small velocity would correspond to a frequency where little wave power is found. 

\begin{figure}
 \includegraphics[width=15 cm, height=10 cm]{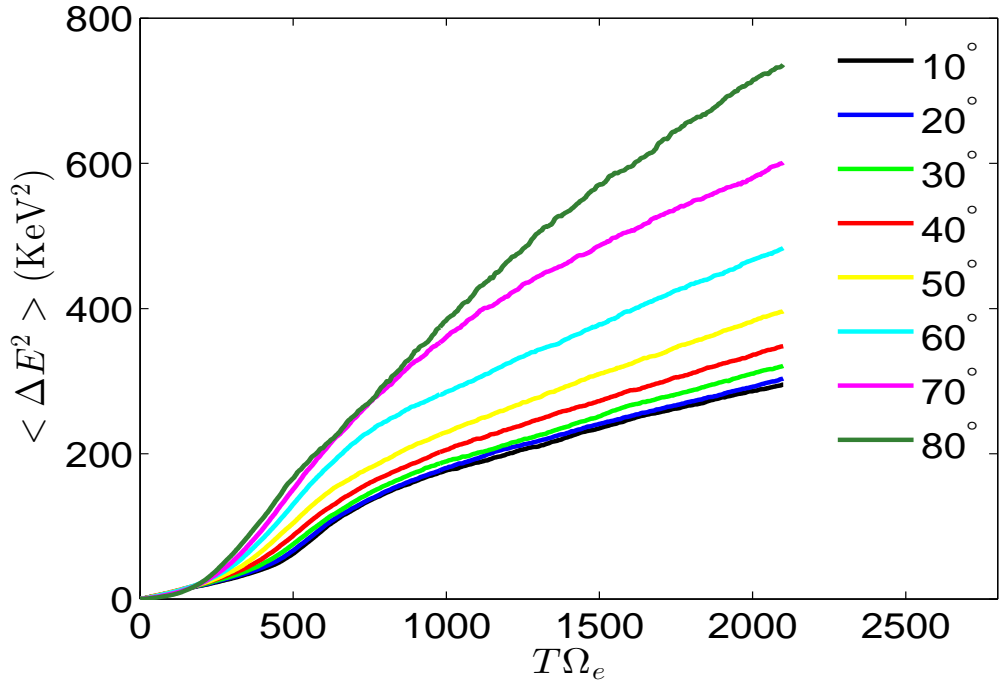}%
 \caption{Evolution of the energy mean squared displacement $\langle \Delta E^2\rangle$ in time for tracked particles. Different colors are for different
 initial pitch angle.}
 \label{fig:variance_E}
 \end{figure}

\begin{figure}
 \includegraphics[width=15 cm, height=10 cm]{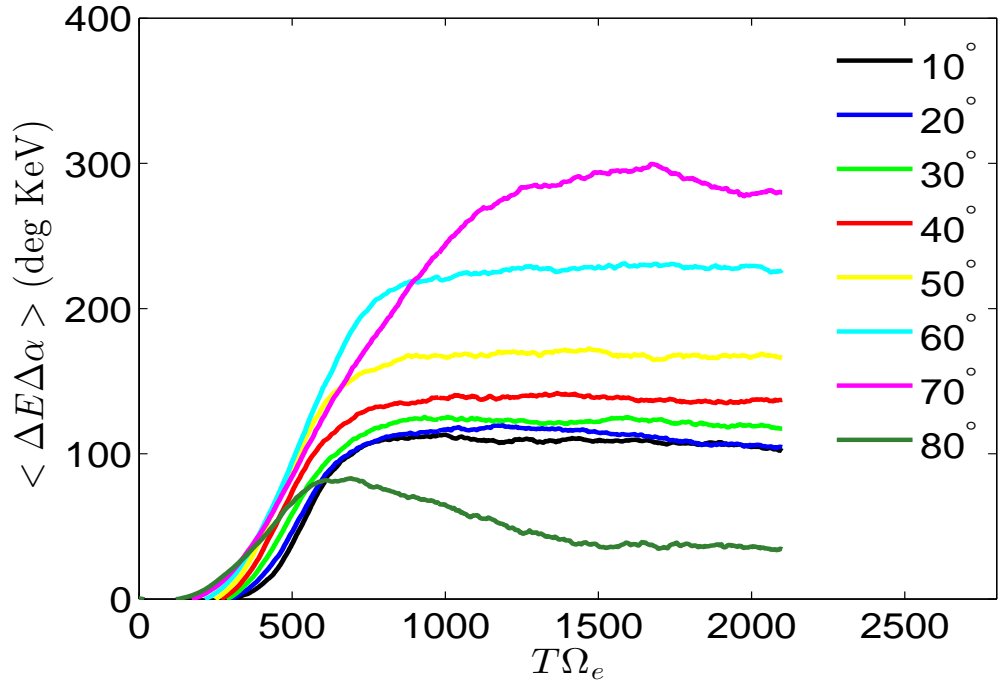}%
 \caption{Evolution of the mean squared displacement $\langle \Delta E\Delta\alpha\rangle$ (i.e. the mixed diffusion term) in time for tracked particles. Different colors are for different
 initial pitch angle.}
 \label{fig:variance_E_pa}
 \end{figure}

\begin{figure}
 \includegraphics[width=15 cm, height=10 cm]{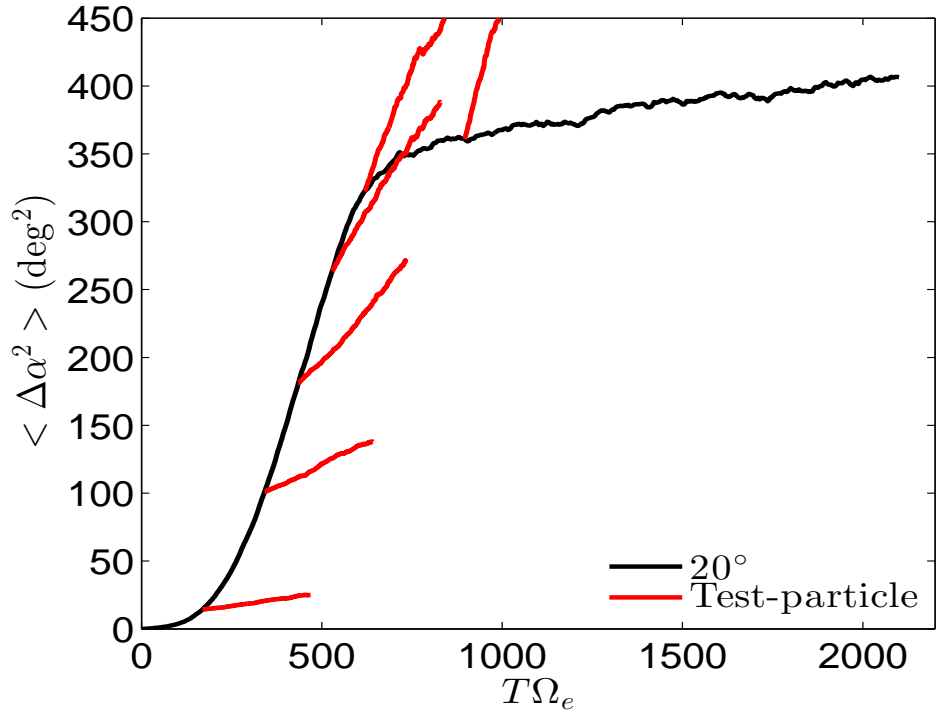}%
 \caption{Comparison of PIC results with test-particle simulations. The black line denotes $\langle \Delta \alpha^2\rangle$ for tracked particles with initial pitch angle $\alpha=20^\circ$ (same as in Figure \ref{fig:variance_pa}). The red lines denotes the results from test-particle. Different red lines 
 are for different simulations initialized with increasing values of $\delta B/B_0$. The lines are then superposed starting from the time 
 at which the same value of $\delta B/B_0$ is reached in the PIC simulation.}
 \label{fig:20_deg_res}
 \end{figure}

\begin{figure}
 \includegraphics[width=15 cm, height=10 cm]{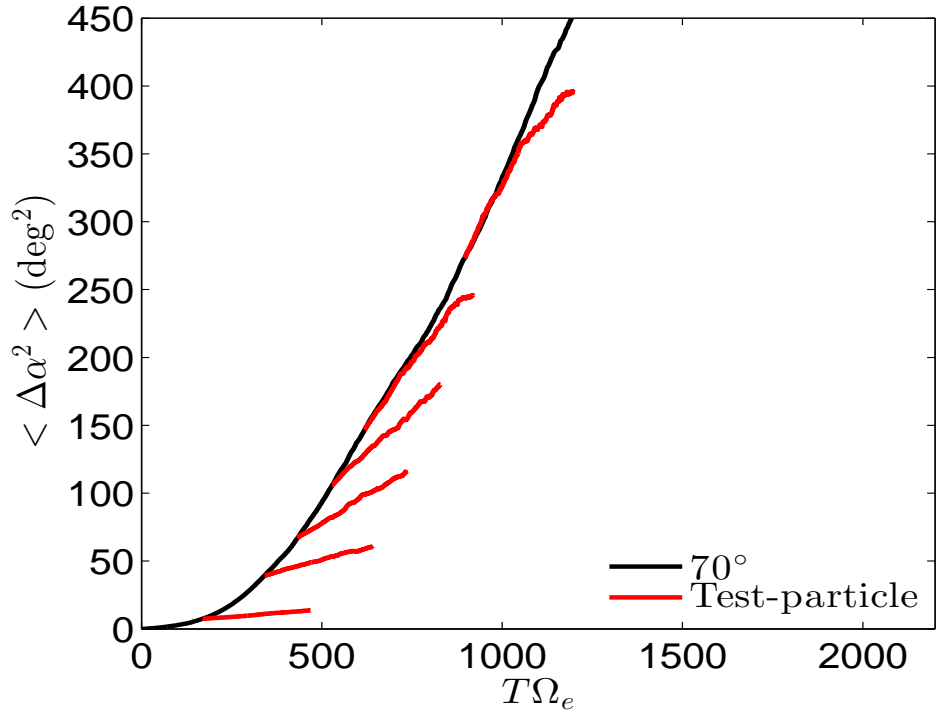}%
 \caption{Same as figure \ref{fig:20_deg_res}, but for initial $\alpha=70^\circ$.}
 \label{fig:70_deg_res}
 \end{figure}

The obtained results are reminiscent of the $90^{\circ}$ scattering problem found by quasi-linear theory for the pitch angle diffusion of cosmic rays \cite[e.g.,][]{goldstein76,qin14} (and many others). Quasi-linear theory can be seen as a first order perturbation theory where the actual particle trajectories are replaced by trajectories in the unperturbed field; this approach however does not allow to correctly describe pitch angle diffusion close to $90^{\circ}$. The development of a nonlinear theory \cite{goldstein76} shows that pitch angle diffusion is indeed very small, but not null, for $\alpha = 90^{\circ}$ and $\delta B/B_0 = 0.05$--$0.1$, while for $\delta B/B_0\simeq 0.3$ the pitch angle scattering rate at $\alpha = 90^{\circ}$ is comparable to that at $\alpha = 60^{\circ}$ (see Figures 1--4 in \cite{goldstein76}). Recently, the scaling of the pitch angle diffusion coefficient with $\delta B/B_0$ and with the cosmic ray energy was considered by \cite{qin14} using both a second order theory and test particle simulations, and they confirmed the smallness of $D_{\alpha \alpha}$ for small to moderate levels of $\delta B/B_0$. Therefore, we can also interpret our results in terms of the nonlinear theories, considering that from Figure 1 for $\Omega_e t > 900$ we have $\delta B/B_0 \simeq 0.06$--$0.07$, corresponding to the range where \cite{goldstein76} found very small scattering.\\ 
For completeness, we show in Figures \ref{fig:variance_E} and \ref{fig:variance_E_pa}
the energy mean squared displacement $\langle\Delta E^2\rangle$,
and the mixed term $\langle\Delta E \Delta \alpha\rangle$, respectively. The role of the mixed diffusion 
coefficient has been recently discussed at length in the literature (see, e.g., \cite{subbotin10,zheng11}),
and Figure \ref{fig:variance_E_pa} confirms that its magnitude is comparable to 
$\langle\Delta \alpha^2\rangle$ and $\langle\Delta E^2\rangle$.\\
To conclude this section we present in Figure \ref{fig:20_deg_res} and \ref{fig:70_deg_res} a comparison between PIC
and test-particle simulations. We interpret test-particle results as representative 
of the quasi-linear paradigm employed in radiation belt simulations, { that is assuming a stationary superposition of waves.}
The aim of such comparison is to show that not taking into account the growth rate of a wave due to an ongoing
kinetic instability, in the calculation of the diffusion coefficients, can lead to an erroneous prediction of pitch angle scattering.
In comparing PIC and test-particle simulations, it is important to remind that in the latter the field amplitude 
is constant, and thus one would not expect a good agreement for long times. Hence the test-particle simulations
are run for 300 gyroperiods only. An important consideration, however, is that the disagreement with PIC is evident since initial times.
Figure \ref{fig:20_deg_res} shows as a black line the pitch angle mean squared displacement
$\langle\Delta \alpha^2\rangle$ for initial $\alpha=20^\circ$ (same plot as in Figure \ref{fig:variance_pa}).
We have superposed, with red lines, the results of test-particle runs. { Each run considers the instantaneous value of magnetic field perturbation $\delta B/B_0$, at that time, to calculate the wave distribution (which does not change in time).}
For clarity, the red lines (test-particle results) starting points are vertically offset so that they are made coincide with the 
PIC result (black line).
The six red lines are for values $\delta B/B_0 = 0.01,0.02,\ldots,0.06$. As expected,
larger values of $\delta B/B_0$ result in a more rapid growth of the mean squared displacement $\langle\Delta \alpha^2\rangle$, for test-particle.
Indeed, if we assume quasi-linear theory to hold, the diffusion coefficient can be calculated as the time derivative
of $\langle\Delta \alpha^2\rangle$, i.e. the slope of the red lines in Figure \ref{fig:20_deg_res}.
As we said, such diffusion coefficient scales quadratically with $\delta B/B_0$ (see Eq. 36 in \cite{summers05}).
A very clear and striking result from Figure \ref{fig:20_deg_res} is that the test-particle prediction underestimates the pitch angle 
scattering in the linear growth phase ($T\Omega_e\lesssim 600$), and largely overestimates the scattering in the 
saturation regime (T$\Omega_e \gtrsim$ 700). The disagreement, which occurs already at initial times, is due to the lack of wave growth, in the test-particle simulations.
Notice that the black line in Figure \ref{fig:20_deg_res} is generated tracking particles that are exactly collimated around $20^\circ$ 
only at initial times, while the test-particle (red lines) always start as exactly collimated. This is a slight inconsistency. However, we have verified that the result does not qualitatively change if one would resample the PIC particles at each time, tracking only the ones that are close to  $20^\circ$ (and with resonant velocity), when the red lines start. The reason can be understood by noticing that the mean square displacement is a weak function of the initial pitch angle (as shown in Figure \ref{fig:variance_pa}), at least for $\alpha\leq60^\circ$.

The same discrepancy shown in Figure \ref{fig:20_deg_res} occurs for all particles with initial $\alpha\lesssim 60^\circ$, that is
particles for which the mean squared displacement in pitch angle 'saturates' in time (Figure \ref{fig:variance_pa}).
We have already commented on the fact that such saturation occurs as the result
that the particles see a strong diffusion barrier at $90^\circ$, and they effectively reach a stationary (or quasi-stationary)
distribution. Of course the diffusion coefficient is very small but not exactly null at $90^\circ$,
and after a sufficiently long time they will diffuse to $\alpha>90^\circ$. 
Such long time evolution is not of interest for radiation belts, since electron precipitation into the loss cone will modify $f(\alpha)$ much earlier. In different contexts, it is important to point out that when electrons are unable to overcome the $90^{\circ}$ barrier, their parallel velocity has a constant sign. This fact gives rise to very long displacements along the magnetic field, which are eventually reversed when $\alpha>90^{\circ}$. When considering spatial diffusion, these long displacements can be at the origin of superdiffusive transport in the parallel direction, as observed in the solar wind \cite[e.g.,][]{perri07} and as discussed by \cite{perrone13,zimbardo13}. Indeed, the $90^{\circ}$ barrier for pitch angle scattering creates a persistent statistical process for $v_{\parallel}$. This also highlights the need to study pitch angle scattering in the nonlinear, self-consistent regime. \\
As expected, a different phenomenology occurs for particles that do not saturate, i.e. for initial pitch angle $\alpha>60^\circ$.
For instance, the case with initial $\alpha=70^\circ$ is plotted in Figure \ref{fig:70_deg_res}.
Here, there seems to be a much better agreement between PIC and test-particle.
However, it is important to notice that, for times $T\lesssim 600$ (i.e. linear growth phase), the test-particle 
still underestimate the pitch angle scattering. The better agreement from time $T \gtrsim 600$ with respect to the
$\alpha=20^\circ$ case (Figure \ref{fig:20_deg_res}) is due to the fact that the magnetic field perturbation
becomes close to saturation and hence the (instantaneous) diffusion coefficient does not vary.
Furthermore, the $70^\circ$ particles are not subject to the $90^\circ$ diffusion barrier, and hence they continue
diffusing. Their mean squared displacement $\langle\Delta\alpha^2 \rangle$ does not saturate abruptly as for the $\alpha=20^\circ$
case and hence there is a more prolonged time for which PIC and test-particle are in an approximate agreement.\
In conclusions, the results for resonant particles (i.e., particles whose initial pitch angle and energy satisfy the resonance
condition with a wave with frequency $\omega/\Omega_e=0.3$) can be summarized as follows.
A distinguishing feature that marks a qualitatively different dynamics is whether the particles diffuse or not through
the $90^\circ$ barrier (within a short timescale).
Particles that do not diffuse through the barrier tend to reach a quasi-stationary isotropic distribution
that fills half domain in pitch angle.
The evolution of their mean squared displacement is strongly correlated with the linear growth and non-linear saturation of the magnetic
field perturbation. Quasi-linear or test-particle predictions does not seem to be applicable to such particles, on the basis
that the ongoing wave growth makes the instantaneous diffusion coefficient (if one still wants to interpret
the dynamics as diffusive) not monotonically correlated with the wave amplitude.
In other words, the time dependency of the wave power and the pitch angle scattering nonlinearly regulate each other.
The crucial point is that most of the scattering occurs during the linear phase, contrary to the assumptions of the resonant limit
of quasi-linear theory previously discussed.
\section{Conclusions}
We have presented PIC and test-particle simulations of resonant wave-particle interactions between lower band whistler modes and  anisotropic electrons, with parameters that realistically mimic the injection of energetic particles at equatorial latitude, for $L\sim 4.3$.
In PIC simulations, the whistler waves are generated self-consistently, and as a consequence the initial particle anisotropy 
is reduced towards a marginal stable configuration. The focus has been on analyzing the statistics of PIC particles, in particular their mean squared displacement in energy and pitch angle, both during the linear growth phase, and in the nonlinear saturation regime. This approach differs from the quasi-linear theory and test-particle simulations, which usually (although not by construction) assume a constant (non-growing) wave field amplitude. We have used test-particle simulations to compare and appreciate the deficiencies of the (resonant limit of the) quasi-linear treatment.
The main results of the paper can be summarized as follows:
\begin{itemize}
 \item The evolution of the mean squared displacements is very well correlated with the linear wave growth and its subsequent saturation.
 Enhanced diffusion is observed during the linear growth phase, in a much larger measure than after saturation. { However, not all pitch angles saturate equally, indicating the importance of nonlinear effects};
 \item For most angles, the distribution in pitch angle saturates and very rapidly reaches a quasi-stationary equilibrium, in a few hundreds gyroperiods, that is in a fraction of the bounce period. { This calls into question the widely used bounce average performed in most radiation belt diffusion calculations};
 \item Although the $90^\circ$ barrier is very effective for most energy/angles, a non-negligible fraction of particles can actually diffuse through the barrier; whether particles diffuse or not through $90^\circ$ determine the dynamics and the saturation (or lack of it) of the mean squared displacement (within the simulation time: because the domain is bounded all particle will eventually saturate in pitch angle);
 \item The disagreement with quasi-linear theory and test-particle simulations can be attributed both on neglecting the rapid growth rate of the linear wave, and on the lack of $90^\circ$ diffusion \citep{camporeale15}.
\end{itemize}
In conclusion, this paper emphasizes the importance of a self-consistent treatment of pitch angle and energy diffusion during the growth phase of whistlers. In a realistic scenario one can envision that the effect of several injection of anisotropic energetic particles in a short time
can result in an overall enhanced diffusion that can cumulatively affect the dynamics of particle loss, and thus should be taken into account 
for realistic estimates.
The inclusion of nonlinear (or higher-order) effect in the calculation of wave-particle interactions is recently becoming a topic of interest, following the discovery of very large amplitude whistler-mode waves in Earth's radiation belts by \cite{cattell08} (see also \cite{kersten11,mozer13}).\\
The results discussed in this paper might also be relevant to other context.
For instance, the generation of suprathermal electrons by resonant wave-particle interactions has been discussed at length for the solar wind \cite[e.g.][]{pierrard99,vocks03,vocks05, saito07}. 
On the other hand, it is well-known that magnetized plasma turbulence exhibits features typical of super or sub-diffusive processes \cite{zimbardo13,perrone13}.
Also, the role of whistler wave is been currently investigated in solar wind turbulence 
\cite{gary09,camporeale11, lacombe14}. Finally, as already mentioned, wave-particle interactions has been a long-time topic well studied in connection to cosmic-ray acceleration \cite{schlickeiser10}.\\
Although this paper has focused on one-dimensional simulations, the implicit PIC algorithm will allow in the near future to tackle fully consistent simulations of wave-particle interaction on multi-dimensions, possibly including multiscale dynamics.

%
%
%
%
%
%
%

\begin{acknowledgments}
We thank X. Tao for sharing his test-particle code, and for insightful comments.
EC wishes to acknowledge J. Bortnik and G. Reeves for useful discussions.
The output data from our simulations is available upon request to the corresponding author (e.camporeale@cwi.nl)
\end{acknowledgments}

%


\newpage
\begin{table}
 \centering
  \caption{Initial energy in KeV for the tracked resonant groups of particles}
  \begin{tabular}{cc}
  \hline
   Initial pitch angle     &  Energy \\
 \hline
 $\alpha=10^\circ$ & 28.5 \\
 $\alpha=20^\circ$ & 31.0 \\
 $\alpha=30^\circ$ & 36.0 \\
 $\alpha=40^\circ$ & 44.9 \\
 $\alpha=50^\circ$ & 61.1 \\
 $\alpha=60^\circ$ & 92.8 \\
 $\alpha=70^\circ$ & 163 \\
 $\alpha=80^\circ$ & 358 \\
 \hline
\end{tabular}
\end{table}

%
%
%
%
%
%
%
%


\end{document}